# The state of health of the Russian population during the pandemic (according to sample surveys)

Leysan Anvarovna Davletshina, Natalia Alekseevna Sadovnikova, Alexander Valeryevich Bezrukov, Olga Guryevna Lebedinskaya
Plekhanov Russian University of Economics

Since the end of 2019, the world has faced a new challenge – an infection that can be both asymptomatic and very severe, spreading rapidly due to the lack of primary immunity in the population.

During this time, significant transformations have taken place in various spheres of activity in the world as a whole, and in Russia in particular – government bodies, businesses, employers of various levels, educational institutions and others were forced to build a certain adaptive model for the realities that almost every inhabitant of the Earth faced as a result. The presence of tangible changes in everyday life, which has already become habitual, the presence of threats to life, health and well-being raises the question of the social well-being of the population in such conditions, which are associated with uncertainty [1].

An integral indicator of the economic development of a state is the health of its population. The preservation and improvement of human health is one of the priorities in the policy of most countries. [4] The difficult epidemiological situation due to the spread of coronavirus infection both in Russia and around the world, the introduction of restrictions on the movement of the population on the territory of the country, as a result, the development of various kinds of difficulties (both economic and social) in most sectors of the economy, together can create a depressing effect on both the economy and the population.

The negative aspects that every resident of Russia faced to one degree or another in 2020 challenge our everyday life, measured everyday life, because at the top, the needs for their own health and the health of their loved ones, the preservation of work and family support are formed [3]. The realities presented by the authors, formed at the junction of 2019-2020, arouse a wide scientific and practical interest in the challenges faced by the inhabitants of Russia and form the relevance of this article.

The quality of statistical information also depends on the method of obtaining data. Today, in the system of state statistics, along with a continuous census of economic entities and the population, their sample surveys conducted in the inter-census period are also practiced [4]. In order to clarify the social well-being of Russian residents, the article will use data from sample surveys as the most digestible assessment tool.

In the article, based on the data of sample observations, a comparative analysis of the social well-being of the population before and after the lockdown was carried out, the comparison was supplemented by the results of the All-Russian survey.

Analysis of the results of the All-Russian survey conducted by VTsIOM and assessment of the relationships

The study formulated the task of identifying the relationship between the respondents ' belonging to a particular socio-demographic group and their social well-being during the period of self-isolation, quarantine or other restrictions imposed during the coronavirus pandemic in 2020 in most regions of the Russian Federation.

On September 27, 2020, a nationwide population survey was conducted, including 11.2 thousand respondents, of whom 45% were men, 55% were women.

The survey method is a formalized telephone interview of the omnibus type. The stratified two-base random sample is based on a complete list of fixed and mobile phone numbers involved in the territory of the Russian Federation. The data are weighted by the probability of selection and by socio-demographic parameters. For this sample, the maximum error size with a 95% probability does not exceed 2.5%.

All federal districts and 80 regions are represented in the survey. The share of rural residents among the respondents is 18-20%. Due to the fact that there are 11 time zones in Russia, the sample was stratified taking into account the time difference. The sample is divided into 5 strata (Center, Urals, Siberia, East, Capitals), within which you can organize the total dial-up time.

To build a sample, all ranges of landline and mobile phone numbers from the Rossvyaz website were used. The total capacity of these ranges is more than 650 million phone numbers — it includes absolutely all phone numbers that are already used by operators or that can be entered according to the current rules. For the survey by the random number sensor, 45 thousand numbers were selected from this database. The phone numbers for ringing are also selected from it by a random number sensor.

The representativeness of the data is ensured by the equal probability of getting into the sample of all Russians over the age of 18, regardless of their place of residence.

In theory, only those Russians who have neither a mobile nor a landline phone are excluded from the sample. According to research, there are less than 1% of them.

Systematic error is reduced due to repeated calls and bounce conversion. Another problem in organizing a sample survey is the refusal of the population (mainly young families) to take part in the survey [5].

Compliance with the technology of random selection theoretically gives a distribution of the main features close to their distribution in the general population.

For the survey, the reachability indicator was calculated using the AAPOR technology (Standard definitions, revised 2016).

During the survey, respondents were asked questions that give a socio-demographic characteristic of the respondents and questions concerning the most important parameters of everyday life, as well as questions that give an idea of the respondents' life values.

Most of the respondents (52.6%) are over 45 years old. Almost half of the respondents (45.2%) have incomplete or higher education. A quarter (24.1%) live in rural areas. From the point of view of the territorial distribution of respondents, 47.6% are residents of the Central and Volga Federal Districts of the Russian Federation.

The most common types of employment of respondents are the following categories: "Unemployed pensioner (including disability)", "Specialist with higher education in the commercial sector", "Skilled worker, including agriculture", which in total corresponds to 54.8% of the total number of respondents. Of the 6,075 respondents, 58.4% work in the commercial sphere, and 35.5% work in the budget sphere.

The analysis of the distribution of income of the population revealed the following structure: the largest group has an income of more than 15 thousand rubles per 1 family member (43.2%). 61.1% of respondents assess the current financial situation as average. At the same time, more than a third (38.4%), providing themselves with basic necessities ("food and clothing"), cannot afford more significant expenses, such as "buying a refrigerator, TV, furniture".

Due to the situation with the coronavirus, a regime of self-isolation, quarantine, or other restrictions was introduced in many regions of Russia. In conditions of limited movement and narrowing of the circle of communication, the risk of contracting a coronavirus infection, immediate illness and the appearance of negative consequences of the disease, the opinion that the inhabitants of the country began to appreciate more during this period becomes relevant.

According to the survey results (an open question, up to three answers), the highest value (24%) corresponds to the option "I find it difficult to answer". The next most common answers are "Your health/relatives" (18%), 12% for the answer options "Family/relatives/friends/Relationships with relatives" and "Nothing", 10% each - "Communication/contacts with people/relatives" and " Work/opportunity to work/Availability of work". Thus, if in some cases the respondents find it difficult to give an answer or do not plan to change anything, the following answers give an understanding of the value of their health and the health of their loved ones, their well-being, as well as the need for communication and work.

For a more detailed disclosure of the respondents ' opinions and the structure of the answers from the point of view of the socio-demographic characteristics of the respondents, as well as an assessment of the differentiation of the answers, we will consider the following questions: "The state of health of your family members", "Family relations", "The opportunity to communicate with friends, relatives", "The opportunity to fully spend leisure time (including rest during vacation)", "The financial situation of your family", "The opportunity to achieve your goals".

According to the results of the survey, concern about the health of family members is very high for both men and women (94% of respondents), in the age ranges of 18-24 years and 25-37 years, the highest values are (97%).

The analysis of respondents by the level of education showed that those who have incomplete secondary education or secondary education (school or vocational school) are the least concerned about the health of family members. Only these groups collectively contain the answers" It doesn't matter at all " (2% each). The same respondents who have a specialized secondary education (technical school) or an incomplete higher education (from the 3rd year of higher education) have a higher education for 95% of the answers in the section "Very important" and 5% "Rather important".

To determine the closeness of the relationship between the respondents ' answers to the question and their gender or age distribution, the coefficients of mutual conjugacy and rank correlation coefficients were determined and analyzed.

When answering the question: "Please tell me, how important are the following aspects of your life for you? The health status of your family members (closed question, one answer)" the distribution of answers does not depend on gender: the opinions of both men and women are absolutely the same ($\rho_{xy} = 1$, $\tau = 1$).

Formulation of hypotheses. H0 – the absence of a relationship between the signs; H1 – an alternative hypothesis, that is, there is a connection between the answers to the question and the age of the respondents.

At a 5% significance level and the number of degrees of freedom of 12 according to the Pearson Distribution, $\chi^2_{tabl} = 21.03$, $\chi^2_{rach} = 108.01$, hence $\chi^2_{rach} > \chi^2_{tabl}$, an alternative hypothesis is confirmed – there is a connection between the answers to the question and the age of the respondents.

Thus, within the framework of the conducted all-Russian survey regarding the question of the health status of family members, a significant part of the respondents consider this aspect to be a very important aspect in their life. At the same time, there are small differences in the responses for various socio-demographic parameters of the respondents.

The family is a kind of "forge of personnel" for society - the main education, the young generation receives value guidelines from their older relatives [6]. In the conditions of restrictions on movement, self-isolation, which took place in most of the regions of the Russian Federation in 2020. there was a situation in which almost all family members were in a limited area (online training, remote work, self-isolation, etc.), which on the one hand made it possible to communicate more and be closer to each other, and created certain difficulties of a domestic, social and other nature.

When answering the question about the importance of relationships in the family, the majority of women (90%) gave the answer "Very important", while for men this answer is 84%, which is quite natural – from time immemorial, a woman has been the keeper of the hearth, collecting and protecting family comfort and comfort.

The analysis of the age distribution when answering this question also presents a fairly logical picture: the largest percentage of "Very important" answers falls on the share of respondents in the age range of 25-34 years (90%), who are most likely young parents whose life guidelines are aimed at raising a child. Respondents in the age range of 45-59 years (potential grandparents) are slightly behind with a value of 89%, which can mean both the continuity of generations and characterize family relations as very important. 87% of responses are in the intervals of 35-44 years and 60 years and older, 77% of responses are in the interval of 18-24 years. 1% of respondents aged 25-34 and 35-44 consider it absolutely not important. A higher value in the range of 60 years and older is 2%. In general, family relations, according to the results of the survey, have a high degree of importance, despite the difficulties that families could face in the conditions of the pandemic.

To determine the closeness of the relationship between the respondents ' answers to the question and their gender or age distribution, the coefficients of mutual conjugacy and rank correlation coefficients were determined and analyzed.

Assessment of the closeness of the connection of the answer to the question: "Please tell me, how important are the following aspects of your life for you? Family relations (closed question, one answer)" when using Spearman's rank correlation coefficient, the dependence of the answers on the gender of the respondents was revealed-0.60 (the relationship is moderate).

Assessing the closeness of the connection of the answer to the question: "Please tell me, how important are the following aspects of your life for you? Family relationships (closed question, one answer)" and the gender of the respondents, the value of the Kendall rank correlation coefficient confirmed their weak dependence (0.40).

Formulation of hypotheses. H0 – the absence of a relationship between the signs; H1 – an alternative hypothesis, that is, there is a connection between the answers to the question and the age of the respondents.

At a 5% significance level and the number of degrees of freedom of 16, based on the Pearson distribution $\chi^2_{tabl} = 26.3$, $\chi^2_{rach} = 295.92$, hence $\chi^2_{rach} > \chi^2_{tabl}$, an alternative hypothesis is confirmed – there is a connection between the answers to the question and the age of the respondents.

Thus, when answering the question about the relationship in the family according to the studied population, most of the answers are characterized as "Very important". The answers "Rather important" are much less common. And the answers "Rather not important" and "Absolutely not important" are insignificant in their scale. It is noteworthy that there is a differentiation of answers to the question posed both in age and territorial sections.

In this paper, the authors analyzed the social well-being of the Russian population during the pandemic on the basis of sample survey data - based on the results of the All-Russian survey conducted by VTsIOM on September 27, 2020.

11.2 thousand people took part in the All-Russian survey conducted by VTsIOM on September 27, 2020. Most of the respondents are of working age, have incomplete or higher education, live in the Central and Volga Federal Districts of Russia, work in the commercial sphere, have an income of more than 15 thousand rubles per 1 family member and assess the current financial situation as average.

The assessment of the relationship between the respondents ' belonging to a particular socio-demographic group and the answers to the question about the health status of family members and the attitude in the family revealed the presence of a significant connection. So, in the gender context, the answers are identical, and in the distribution into age groups, an alternative hypothesis is confirmed.